\RequirePackage{lineno}
\documentclass[aps,prl,twocolumn,showpacs,superscriptaddress,groupedaddress]{revtex4}  
\usepackage{graphicx}  
\usepackage{dcolumn}   
\usepackage{bm}        
\usepackage{amssymb}   
\usepackage{epsfig}
\usepackage{slashed}
\usepackage{multirow}
\usepackage{dcolumn}
\usepackage{subfigure}

\usepackage{tabularx}
\newcolumntype{C}{>{\centering}X}

\begin{document}

\widetext

\hspace{5.2in} \mbox{FERMILAB-PUB-11-616-E}

\title{$\boldsymbol{Z\gamma}$ production and limits on anomalous $\boldsymbol{ZZ\gamma}$ and $\boldsymbol{Z\gamma\gamma}$ couplings in $\boldsymbol{p\bar{p}}$ collisions at $\boldsymbol{\sqrt{s}=1.96}$ TeV }
%
\affiliation{Universidad de Buenos Aires, Buenos Aires, Argentina}
\affiliation{LAFEX, Centro Brasileiro de Pesquisas F{\'\i}sicas, Rio de Janeiro, Brazil}
\affiliation{Universidade do Estado do Rio de Janeiro, Rio de Janeiro, Brazil}
\affiliation{Universidade Federal do ABC, Santo Andr\'e, Brazil}
\affiliation{Instituto de F\'{\i}sica Te\'orica, Universidade Estadual Paulista, S\~ao Paulo, Brazil}
\affiliation{University of Science and Technology of China, Hefei, People's Republic of China}
\affiliation{Universidad de los Andes, Bogot\'{a}, Colombia}
\affiliation{Charles University, Faculty of Mathematics and Physics, Center for Particle Physics, Prague, Czech Republic}
\affiliation{Czech Technical University in Prague, Prague, Czech Republic}
\affiliation{Center for Particle Physics, Institute of Physics, Academy of Sciences of the Czech Republic, Prague, Czech Republic}
\affiliation{Universidad San Francisco de Quito, Quito, Ecuador}
\affiliation{LPC, Universit\'e Blaise Pascal, CNRS/IN2P3, Clermont, France}
\affiliation{LPSC, Universit\'e Joseph Fourier Grenoble 1, CNRS/IN2P3, Institut National Polytechnique de Grenoble, Grenoble, France}
\affiliation{CPPM, Aix-Marseille Universit\'e, CNRS/IN2P3, Marseille, France}
\affiliation{LAL, Universit\'e Paris-Sud, CNRS/IN2P3, Orsay, France}
\affiliation{LPNHE, Universit\'es Paris VI and VII, CNRS/IN2P3, Paris, France}
\affiliation{CEA, Irfu, SPP, Saclay, France}
\affiliation{IPHC, Universit\'e de Strasbourg, CNRS/IN2P3, Strasbourg, France}
\affiliation{IPNL, Universit\'e Lyon 1, CNRS/IN2P3, Villeurbanne, France and Universit\'e de Lyon, Lyon, France}
\affiliation{III. Physikalisches Institut A, RWTH Aachen University, Aachen, Germany}
\affiliation{Physikalisches Institut, Universit{\"a}t Freiburg, Freiburg, Germany}
\affiliation{II. Physikalisches Institut, Georg-August-Universit{\"a}t G\"ottingen, G\"ottingen, Germany}
\affiliation{Institut f{\"u}r Physik, Universit{\"a}t Mainz, Mainz, Germany}
\affiliation{Ludwig-Maximilians-Universit{\"a}t M{\"u}nchen, M{\"u}nchen, Germany}
\affiliation{Fachbereich Physik, Bergische Universit{\"a}t Wuppertal, Wuppertal, Germany}
\affiliation{Panjab University, Chandigarh, India}
\affiliation{Delhi University, Delhi, India}
\affiliation{Tata Institute of Fundamental Research, Mumbai, India}
\affiliation{University College Dublin, Dublin, Ireland}
\affiliation{Korea Detector Laboratory, Korea University, Seoul, Korea}
\affiliation{CINVESTAV, Mexico City, Mexico}
\affiliation{Nikhef, Science Park, Amsterdam, the Netherlands}
\affiliation{Radboud University Nijmegen, Nijmegen, the Netherlands and Nikhef, Science Park, Amsterdam, the Netherlands}
\affiliation{Joint Institute for Nuclear Research, Dubna, Russia}
\affiliation{Institute for Theoretical and Experimental Physics, Moscow, Russia}
\affiliation{Moscow State University, Moscow, Russia}
\affiliation{Institute for High Energy Physics, Protvino, Russia}
\affiliation{Petersburg Nuclear Physics Institute, St. Petersburg, Russia}
\affiliation{Instituci\'{o} Catalana de Recerca i Estudis Avan\c{c}ats (ICREA) and Institut de F\'{i}sica d'Altes Energies (IFAE), Barcelona, Spain}
\affiliation{Stockholm University, Stockholm and Uppsala University, Uppsala, Sweden}
\affiliation{Lancaster University, Lancaster LA1 4YB, United Kingdom}
\affiliation{Imperial College London, London SW7 2AZ, United Kingdom}
\affiliation{The University of Manchester, Manchester M13 9PL, United Kingdom}
\affiliation{University of Arizona, Tucson, Arizona 85721, USA}
\affiliation{University of California Riverside, Riverside, California 92521, USA}
\affiliation{Florida State University, Tallahassee, Florida 32306, USA}
\affiliation{Fermi National Accelerator Laboratory, Batavia, Illinois 60510, USA}
\affiliation{University of Illinois at Chicago, Chicago, Illinois 60607, USA}
\affiliation{Northern Illinois University, DeKalb, Illinois 60115, USA}
\affiliation{Northwestern University, Evanston, Illinois 60208, USA}
\affiliation{Indiana University, Bloomington, Indiana 47405, USA}
\affiliation{Purdue University Calumet, Hammond, Indiana 46323, USA}
\affiliation{University of Notre Dame, Notre Dame, Indiana 46556, USA}
\affiliation{Iowa State University, Ames, Iowa 50011, USA}
\affiliation{University of Kansas, Lawrence, Kansas 66045, USA}
\affiliation{Kansas State University, Manhattan, Kansas 66506, USA}
\affiliation{Louisiana Tech University, Ruston, Louisiana 71272, USA}
\affiliation{Boston University, Boston, Massachusetts 02215, USA}
\affiliation{Northeastern University, Boston, Massachusetts 02115, USA}
\affiliation{University of Michigan, Ann Arbor, Michigan 48109, USA}
\affiliation{Michigan State University, East Lansing, Michigan 48824, USA}
\affiliation{University of Mississippi, University, Mississippi 38677, USA}
\affiliation{University of Nebraska, Lincoln, Nebraska 68588, USA}
\affiliation{Rutgers University, Piscataway, New Jersey 08855, USA}
\affiliation{Princeton University, Princeton, New Jersey 08544, USA}
\affiliation{State University of New York, Buffalo, New York 14260, USA}
\affiliation{Columbia University, New York, New York 10027, USA}
\affiliation{University of Rochester, Rochester, New York 14627, USA}
\affiliation{State University of New York, Stony Brook, New York 11794, USA}
\affiliation{Brookhaven National Laboratory, Upton, New York 11973, USA}
\affiliation{Langston University, Langston, Oklahoma 73050, USA}
\affiliation{University of Oklahoma, Norman, Oklahoma 73019, USA}
\affiliation{Oklahoma State University, Stillwater, Oklahoma 74078, USA}
\affiliation{Brown University, Providence, Rhode Island 02912, USA}
\affiliation{University of Texas, Arlington, Texas 76019, USA}
\affiliation{Southern Methodist University, Dallas, Texas 75275, USA}
\affiliation{Rice University, Houston, Texas 77005, USA}
\affiliation{University of Virginia, Charlottesville, Virginia 22901, USA}
\affiliation{University of Washington, Seattle, Washington 98195, USA}
\author{V.M.~Abazov} \affiliation{Joint Institute for Nuclear Research, Dubna, Russia}
\author{B.~Abbott} \affiliation{University of Oklahoma, Norman, Oklahoma 73019, USA}
\author{B.S.~Acharya} \affiliation{Tata Institute of Fundamental Research, Mumbai, India}
\author{M.~Adams} \affiliation{University of Illinois at Chicago, Chicago, Illinois 60607, USA}
\author{T.~Adams} \affiliation{Florida State University, Tallahassee, Florida 32306, USA}
\author{G.D.~Alexeev} \affiliation{Joint Institute for Nuclear Research, Dubna, Russia}
\author{G.~Alkhazov} \affiliation{Petersburg Nuclear Physics Institute, St. Petersburg, Russia}
\author{A.~Alton$^{a}$} \affiliation{University of Michigan, Ann Arbor, Michigan 48109, USA}
\author{G.~Alverson} \affiliation{Northeastern University, Boston, Massachusetts 02115, USA}
\author{G.A.~Alves} \affiliation{LAFEX, Centro Brasileiro de Pesquisas F{\'\i}sicas, Rio de Janeiro, Brazil}
% ...
%
%
\collaboration{The D0 Collaboration\footnote{with visitors from
$^{a}$Augustana College, Sioux Falls, SD, USA,
$^{b}$The University of Liverpool, Liverpool, UK,
$^{c}$UPIITA-IPN, Mexico City, Mexico,
$^{d}$SLAC, Menlo Park, CA, USA,
$^{e}$University College London, London, UK,
$^{f}$Centro de Investigacion en Computacion - IPN, Mexico City, Mexico,
$^{g}$ECFM, Universidad Autonoma de Sinaloa, Culiac\'an, Mexico,
and 
$^{h}$Universit{\"a}t Bern, Bern, Switzerland.
$^{\ddag}$Deceased.
}} \noaffiliation
\vskip 0.25cm
     
\date{15 November, 2011}

\begin{abstract}
We present a measurement of
$p\bar{p} \rightarrow Z\gamma \rightarrow \ell^+\ell^-\gamma$ ($\ell = e$, $\mu$)
production with a data sample corresponding to an integrated luminosity of 6.2 fb$^{-1}$ 
collected by the D0 detector at the Fermilab Tevatron  $p\bar{p}$ Collider.  
The results of the electron and muon channels are combined, and we measure the total production cross section and the differential cross section $d\sigma/dp_T^\gamma$, where $p_T^\gamma$ is the momentum of the photon in the plane transverse to the beamline. The results obtained are consistent with the standard model predictions from next-to-leading order calculations.  
We use the transverse momentum spectrum of the photon to place limits on  anomalous 
$ZZ\gamma$ and $Z\gamma\gamma$ couplings.
\end{abstract}

\maketitle

\section{Introduction}

\begin{figure}[htbp]
\subfigure[]{\label{fig:SM1}\includegraphics[width=0.2\textwidth]{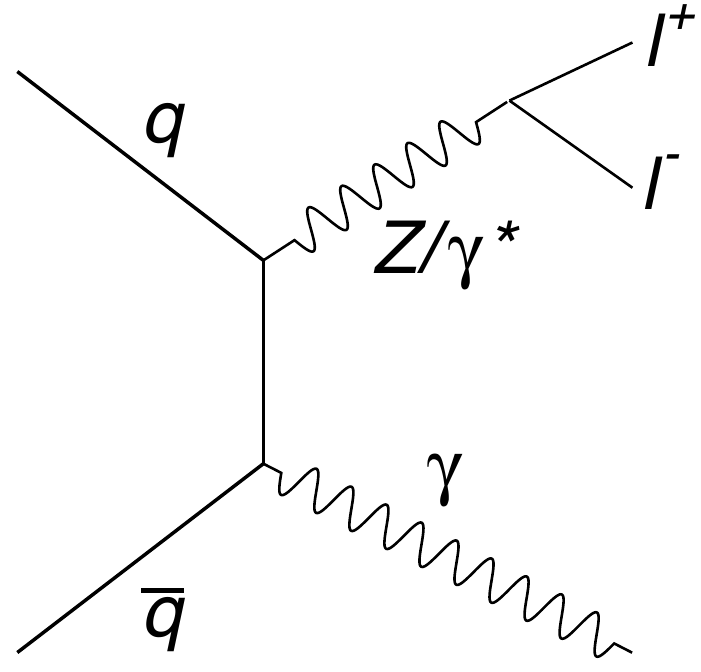}}
\subfigure[]{\label{fig:SM2}\includegraphics[width=0.2\textwidth]{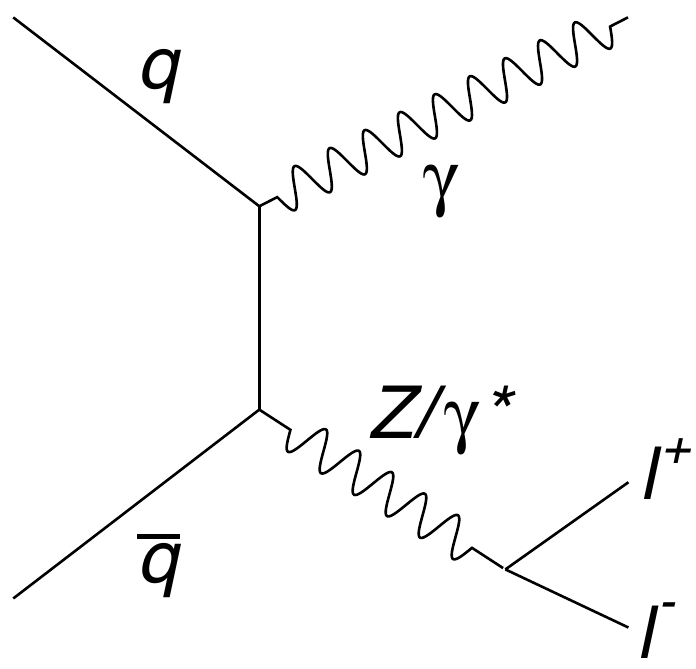}}
\subfigure[]{\label{fig:SM1}\includegraphics[width=0.2\textwidth]{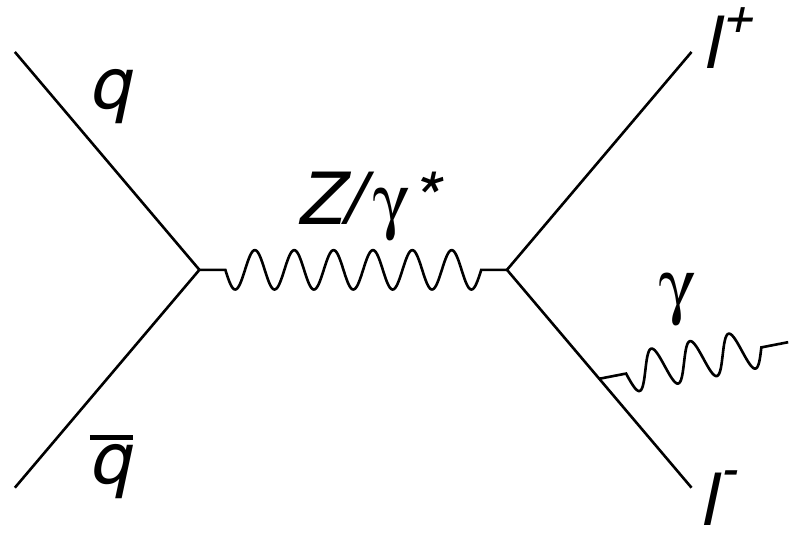}}
\subfigure[]{\label{fig:SM2}\includegraphics[width=0.2\textwidth]{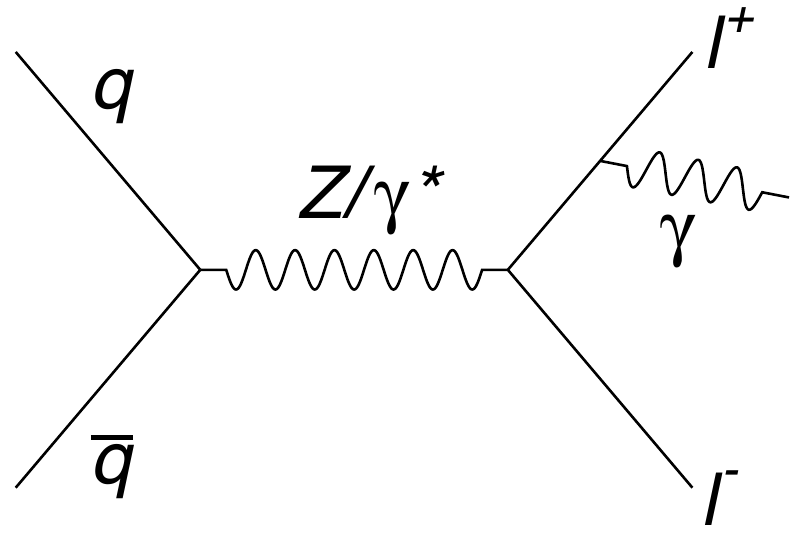}}
\caption{Feynman diagrams for leading-order $Z\gamma$
  production in the SM:
  (a) and (b) initial-state radiation from one of the initial-state partons,
  (c) and (d) final-state radiation from one of the final-state leptons.
 }
\label{ISRFSR}
\end{figure}

The standard model (SM) describes the electroweak interactions through a non-abelian gauge group $SU(2)_L \otimes U(1)_Y$, which includes self-interactions of gauge bosons.  Because the $Z$ boson carries no electric charge, a coupling between a $Z$ boson and a photon is not permitted.  The $Z\gamma$ production in the SM is dominated by the lowest-order Feynman diagrams shown in  Fig.~\ref{ISRFSR}.  
\begin{figure}[htbp]
\subfigure[]{\label{fig:ISR1}\includegraphics[width=0.22\textwidth]{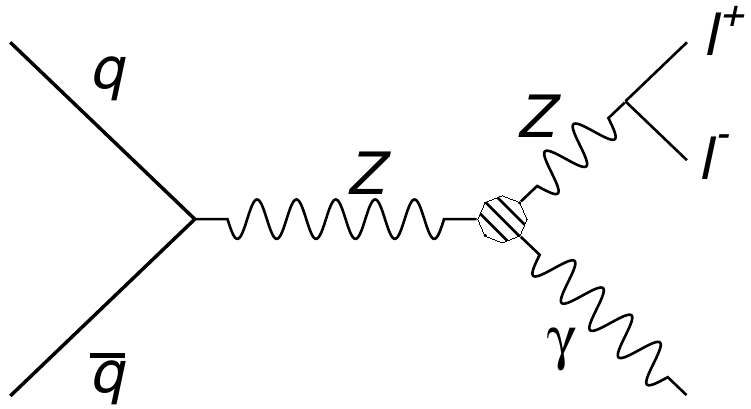}}
\subfigure[]{\label{fig:ISR2}\includegraphics[width=0.22\textwidth]{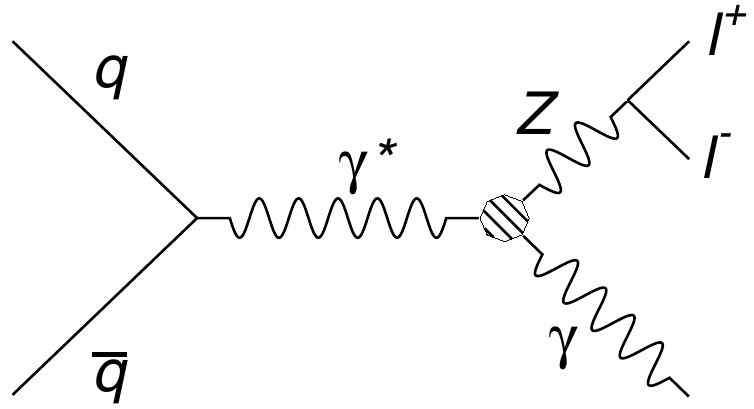}}
\caption{\small
  Feynman diagrams illustrating anomalous $Z\gamma$ production with a  $ZZ\gamma$ vertex (a) and a $Z\gamma\gamma$ vertex (b) .
 }
\label{ACs}
\end{figure}

An excess in the number of high-energy photons can be a sign of new physics, e.g.,~supersymmetry, as described in Ref.~\cite{zgammaSUSY} or new
  heavy fermions with nonstandard couplings to the gauge
  bosons, as discussed in Ref.~\cite{zgammaSUSY2}.  Such an excess of high-energy photons can be described by assuming only  Lorentz and local $U(1)_{em}$ gauge invariant $ZZ\gamma$ and $Z\gamma\gamma$ trilinear gauge boson vertices of the form 
shown in Fig.~\ref{ACs}, using an effective theory with 
eight complex coupling parameters, $h^V_i$, where $i$ = $1,~2,~3,~4$ and $V=Z$ 
or $\gamma$ \cite{hzp}.  Here, the couplings parameters $h^V_1$ and $h^V_3$ ($h^V_2$ and $h^V_4$) are associated with dimension-six (dimension-eight) operators which allow for an interaction between a $Z$ boson and a photon.   
To conserve tree-level unitarity at asymptotically high energies, one can introduce form factors dependent on the square of the partonic center-of-mass energy, $\hat{s}$, given by $h^V_i = h^V_{0i}/(1+\hat{s}/\Lambda^2)^n$, where $\Lambda$ is the mass scale at which the new physics responsible for anomalous couplings
is introduced \cite{baur}.  These anomalous gauge boson couplings would give rise to
an excess of photons at high transverse momentum, $p_T^\gamma$, which can be searched for by measuring the total production cross section and
the differential cross section $d\sigma/dp_T^\gamma$ for $Z\gamma\rightarrow \ell^+\ell^-\gamma$ ($\ell\ell\gamma$ henceforth) production.  If no evidence of new physics is seen, 
we can place limits on the real components of the $CP$-even 
coupling parameters, $h^V_{03}$ and $h^V_{04}$, for $\Lambda = 1.2$ and $1.5$ TeV.   Following Ref.~\cite{baur}, 
we choose form-factor powers for the unitarity scaling dimensions of  $n=3$ for $h^V_3$ and $n=4$ for $h^V_4$.  $Z\gamma$ production has been previously studied at collider experiments \cite{zgammaprev8, d0prev, zgammaprev5, zgammaprev2, zgammaprev7, zgammaprev6, zgammaprev4, zgammaprev3, zgammaprev1}, and because the value of $\Lambda$ greatly affects the scale of anomalous $Z\gamma$ production, we choose to perform this analysis for the values of $\Lambda$ that were used by the recent D0 \cite{zgammaprev8, d0prev} and CDF \cite{zgammaprev5} analyses.  This choice of $\Lambda$ differs from the value used by the ALEPH \cite{zgammaprev2}, CMS \cite{zgammaprev6}, DELPHI \cite{zgammaprev4}, L3 \cite{zgammaprev3}, and OPAL \cite{zgammaprev1} collaborations. 

We present measurements of the 
inclusive cross section and differential 
cross section for $Z\gamma$ production in the electron and muon channels
using a data sample corresponding to an integrated luminosity
of 6.2 $\pm$ 0.4 fb$^{-1}$ collected at 
$\sqrt{s} = 1.96$ TeV by the D0 detector at the Fermilab Tevatron Collider between June 2006 and July 2010.  These results provide a significant improvement 
in the sensitivity to anomalous $ZZ\gamma$ and $Z\gamma\gamma$ production compared to a previous D0 publication, utilizing the same channels and an integrated luminosity of 1 fb$^{-1}$ \cite{zgammaprev8}.  In addition to increasing the size of the data set, we also combine with a previous result in the same channels \cite{zgammaprev8}, along with another D0 result \cite{d0prev} that used 3.6 fb$^{-1}$ of $Z\gamma\rightarrow\nu\nu\gamma$ production to place stringent limits on $Z\gamma$ anomalous couplings.

\section{ The D0 Detector}
The D0 detector \cite{dzero1,dzero2, dzero3, dzero4, dzero5} consists of a central tracking
system contained within a 2~T superconducting solenoidal magnet,
surrounded by a central preshower (CPS) detector, a
liquid-argon sampling calorimeter, and an outer muon system.
The tracking system, consisting of a silicon microstrip tracker (SMT)
and a scintillating fiber tracker (CFT), provides coverage
for charged particles in the pseudorapidity range
$|\eta| \lesssim 3$~\cite{d0_coordinate}.
The CPS is located immediately before the inner layer of the calorimeter
and has about one radiation length of absorber followed by several
layers of scintillating strips.
The calorimeter consists of a central cryostat sector (CC) with
coverage $|\eta| \lesssim 1.1$ and two end calorimeters (EC) which extend coverage to $|\eta| \approx 4.2$.
The electromagnetic (EM) section of the
calorimeter is segmented into four longitudinal layers
(EM$i$, $i=1,~4$) with transverse
segmentation of
$\Delta\eta\times\Delta\phi = 0.1\times 0.1$,
except in EM3,
where it is $0.05\times 0.05$.
The muon system resides beyond the calorimeter and consists of a
layer of tracking detectors and scintillation trigger counters
before a 1.8 T iron toroidal magnet, followed by two
similar layers after the toroid. The coverage of the muon
system corresponds to a pseudorapidity range $|\eta| < 2$.

\section{Event selection}
Candidate $Z\gamma$ events are selected in the $e^+e^-\gamma$ 
and $\mu^+\mu^- \gamma$ ($ee\gamma$ and $\mu\mu\gamma$ henceforth) final states.
The $p\bar{p}$ interaction vertex must be reconstructed within $\pm 60$~cm of
the center of the D0 detector along the beam ($z$) axis.
For the electron channel,
a sample of candidate $Z$-boson events is collected with a suite of single-electron triggers.
The electrons are selected by requiring an EM cluster in either
the CC ($|\eta| < 1.1$) or EC ($1.5 < |\eta| < 2.5$) regions of the EM calorimeter
with transverse momentum $p_T > 25$ (15)~GeV/$c$ for the electron candidate with the highest (next-to-highest) transverse energy 
contained within a cone of radius
$\Delta\mathcal{R}=\sqrt{(\Delta \eta)^2 + (\Delta \phi)^2}=0.2$,
centered on the axis of the EM shower.
At least 90\% of the cluster energy must be deposited within the EM section
of the calorimeter.
Electron candidates, with a shower shape consistent with that
of an electron, are required to be spatially matched to a track and
to be isolated in both the calorimeter and tracking detectors.
To suppress jets and photons misidentified as electrons, a likelihood discriminant
is built using a set of variables sensitive to differences in tracker activity
and energy deposits in the calorimeter:
the number of tracks and the scalar sum of the transverse momentum of all tracks
within $\Delta\mathcal{R} < 0.4$ of the EM cluster,
the fraction of energy deposited in the EM section of the calorimeter,
the longitudinal and transverse shower profile in the calorimeter,
and the ratio between the transverse energy in the calorimeter and the transverse momentum
of the electron associated track.
To further suppress jets misidentified as electrons, in particular, for high instantaneous luminosity conditions, a neural network algorithm
is trained on Drell-Yan $Z/\gamma^* \to e^+e^-$ and jet data, using
information from the calorimeter and CPS:
the numbers of cells above
a threshold in EM1 within $\Delta\mathcal{R} < $ 0.2 and 0.2 $< \Delta\mathcal{R} <$ 0.4
of the EM cluster, the number of CPS clusters within
$\Delta\mathcal{R} < $ 0.1 of the EM cluster, and the squared-energy weighted
width of the energy deposit in the CPS.
Events where both electrons are contained within the EC are excluded because of the small signal acceptance.  Candidate events where the $Z$ boson decays into two muons
are collected using a suite of single-muon triggers.
Within the muon channel, muon candidates are required to be within $|\eta| < 2$ and matched to a well-isolated track in both the tracker and the calorimeter with transverse momentum $p_T > 15$ GeV/$c$.  The highest $p_{T}$ muon must have $p_T$ $>$ 20 GeV/$c$. Both muon candidates are required to originate from within 2 cm of 
the interaction point in the $z$ direction.

Photon candidates in both the electron and muon channels are
required to have transverse momentum $p_T^{\gamma} > 10$ GeV/$c$
within a cone of radius $\Delta\mathcal{R}=0.2$ 
centered around the EM shower in the CC.  The rapidity of
the photon, $\eta^\gamma$, is required to be $|\eta^\gamma|<1.1$.
Additionally, the photon candidate must satisfy the following requirements:
(i) at least $90\%$ of the cluster energy is deposited in the
EM calorimeter; (ii) the calorimeter isolation variable
$I = [E_{\text{tot}}(0.4)-E_{\text{EM}}(0.2)]/E_{\text{EM}}(0.2) <$~0.15, where $E_{\text{tot}}(0.4)$ is the
total energy in a cone of radius $\Delta\mathcal{R}=0.4$ and
$E_{\text{EM}}(0.2)$ is the EM energy in a cone of radius $\Delta\mathcal{R}=0.2$;
(iii) the energy-weighted cluster width
in the EM3 layer is consistent with that for an EM shower;
(iv) the scalar sum of the $p_T$ of all tracks,
$p_{T_{\text{trk}}}^{\text{sum}}$, originating from the
interaction point in an annulus of $0.05<\Delta\mathcal{R}<0.4$
around the cluster
is less than 2.0 GeV/$c$;
(v) the EM cluster must not be spatially matched to
either a reconstructed track or to energy depositions in the SMT or
CFT detectors that are compatible with a trajectory of an electron \cite{HOR};
and (vi) an output larger than 0.1 of an artificial neural network ($O_{NN}$)~\cite{ANN5}
that combines information from a set of variables sensitive to
differences between photons and jets
in the tracking detector, the calorimeter, and the CPS detector.

The dilepton invariant mass, $M_{\ell\ell}$, is required
to be greater than 60 GeV/$c^2$, and the photon must be separated from each lepton
by $\Delta\mathcal{R}_{\ell\gamma} > 0.7$.  Additionally, each lepton must be separated from a jet by $\Delta\mathcal{R}_{\ell j} > 0.5$.  
In the electron and muon channels, we select 1002 and 1000 data events, respectively.  In order to reduce the contribution of final-state radiation (FSR), subset data samples are defined with the requirement that the reconstructed three-body invariant mass, $M_{\ell\ell\gamma}$, exceeds 110 GeV/$c^2$.
With this additional requirement, 304 and 308 data events are selected in the electron
and muon channels, respectively.

\subsection{Background subtraction}
The selected sample is contaminated by a small admixture of $Z+$jet events in which
a jet is misidentified as a photon.
To estimate this background in the electron channel, the fraction of jets that pass the
photon selection criteria but fail either the  $p_{T_{\text{trk}}}^{\text{sum}}$ or the shower width requirement, as determined
by using a dijet data sample, is parametrized as a function of $p_T^{\gamma}$ and $\eta_{\gamma}$ (ratio method). 
The background from $Z$+jet production is then estimated starting from a data sample
obtained by reversing the requirements either on  $p_{T_{\text{trk}}}^{\text{sum}}$ or on the shower width requirement,
and applying the same parametrization.
  A systematic uncertainty associated with the estimation of the number of real photons in the data samples is due to the finite size of the dijet background sample.  
After subtracting the estimated background from the data in the electron channel, we estimate 926 $\pm$ 53 (stat.)~$\pm$ 19 (syst.)~signal events when no $M_{\ell\ell\gamma}$ requirement is applied,
and 255 $\pm$ 15 (stat.)~$\pm$ 5 (syst.)~signal events with $M_{\ell\ell\gamma}>110$ GeV/$c^2$. 

To estimate the background in the muon channel, we use a matrix method to estimate the $Z$+jet background contribution.
After applying all of the selection criteria described above,
a tighter requirement on $O_{NN}$ is used to classify the 
data events into two categories, depending on
whether the photon candidate passes ($p$) or fails ($f$) this requirement.
The corresponding number of events compose a 2-component vector ($N_p$, $N_f$).
Thus, the sample composition is obtained by resolving a linear
system of equations
$(N_{p}, N_{f})^T = {\cal E}\times (N_{Z\gamma}, N_{Zj})^T$,
where $N_{Z\gamma}$ ($N_{Zj}$) is the true number of $Z+\gamma$ ($Z+$jet)
events in the fiducial region. The $2\times2$ efficiency matrix ${\cal E}$ contains
the photon $\varepsilon_{\gamma}$ and jet $\varepsilon_{\rm jet}$ efficiencies that are estimated using photon and jet Monte Carlo (MC) samples and validated in data.
Based on these studies, the efficiencies are parametrized as a function of the photon candidates'
$\eta^\gamma$ with 1.5\% and
10\% relative systematic uncertainties for $\varepsilon_{\gamma}$
and $\varepsilon_{\rm jet}$ respectively.  
Having subtracted the estimated background from data in the muon channel, we estimate 947 $\pm$ 40 (stat.)~$\pm$ 16 (syst.)~signal events when no $M_{\ell\ell\gamma}$ requirement is applied,
and 285 $\pm$ 24 (stat.)~$\pm$ 2 (syst.)~signal events requiring $M_{\ell\ell\gamma}>110$ GeV/$c^2$.

As a cross-check, the $Z+$jet background is also estimated through a fit
to the shape of the $O_{NN}$ distribution in data for both electron and muon channels,
using MC templates constructed from simulated photon and jet events. 
The results are in good agreement with those obtained from
the ratio and matrix methods.

\section{Results}
\subsection{Total cross section}
The total cross section for $\ell\ell\gamma$ production is obtained from the 
ratio of the acceptance-corrected  $\ell\ell\gamma$ rate for 
$M_{\ell\ell} > 60$ GeV/$c^2$, $\Delta \mathcal{R}_{\ell\gamma} > 0.7$, 
$p_T^\gamma > 10$ GeV/$c$, and $|\eta^\gamma| < 1$, 
to the total acceptance-corrected dilepton rate for $M_{\ell\ell} > 60$ GeV/$c^2$.  Henceforth, these acceptance requirements are referred to as the generator-level requirements.  
We utilize this method because uncertainties 
associated with the trigger efficiencies, reconstruction efficiencies, 
and integrated luminosity are larger than the theoretical uncertainties and cancel in the ratio.  This ratio is multiplied by a theoretical estimate for the total cross section for inclusive $Z/\gamma^*\rightarrow \ell\ell$ production 
for $M_{\ell\ell}>60$ GeV/$c^2$:
\begin{equation}
\label{xsect}
\sigma_{Z\gamma} \times \mathcal{B}= \frac{ \kappa N_{\ell\ell\gamma}^{\text{data}}\left(A\times\epsilon_{ID}\right)_{\ell\ell\gamma}^{-1}}{N_{\ell\ell}^{\text{data}}\left(A\times\epsilon_{ID}\right)_{\ell\ell}^{-1}}  \times (\sigma_Z\times \mathcal{B})_{\text{FEWZ}}^{\text{NNLO}}.
\end{equation}
Here, $N_{\ell\ell}^{\text{data}}$ and $N_{\ell\ell\gamma}^{\text{data}}$ are the number of measured $Z$ and background-subtracted $Z\gamma$ events in the data sample, respectively.  The factor $(\sigma_Z\times \mathcal{B})_{\text{FEWZ}}^{\text{NNLO}}$ is calculated with the {\sc fewz} 
next-to-next-to-leading-order (NNLO) generator \cite{fewz1}-\cite{fewz2}, 
with the CTEQ6.6 parton distribution functions (PDF) \cite{CTEQ66}.  
The {\sc fewz} theoretical prediction is 262.9 $\pm$ 8.0 pb, 
where the dominant uncertainty is
from the choice of PDF.    The term $\mathcal{B}$ is the branching fraction for $Z/\gamma^*\rightarrow \ell\ell$,  which in the SM is 3.4\% for either electrons or muons.   
The factor $\kappa$ corrects for the resolution effects that would cause events not to pass the selections on the generator-level quantities, e.g.~a generator-level photon with $p_T^\gamma < 10$ GeV/$c$, but to pass the reconstruction requirements, e.g.~a reconstructed photon with $p_T^\gamma > 10$ GeV/$c$.  This factor is only used for $Z\gamma\rightarrow \ell\ell\gamma$ events, and
corrects for the photon energy smearing that dominates in the first $p_T^\gamma$ bin.  The muon $p_T$ resolution affects both $Z/\gamma^*\rightarrow\ell\ell$ 
and $Z\gamma\rightarrow \ell\ell\gamma$ and the corresponding correction cancel in the ratio of cross sections.  For the events that pass the generator-level requirements, the factors
$(A\times\epsilon_{ID})_{\ell\ell\gamma}$ and $(A\times\epsilon_{ID})_{\ell\ell}$ provide the fraction of events 
that pass the analysis requirements, with all acceptances measured 
relative to the kinematic requirements at the generator level for the $\ell\ell$ and $\ell\ell\gamma$ final states, respectively.   
Events migrate between bins in $p_T^\gamma$ because of finite detector resolution, 
and these effects are taken into account in calculating 
$(A\times\epsilon_{ID})_{\ell\ell\gamma}$ as a function of $p_T^\gamma$, while $(A\times\epsilon_{ID})_{\ell\ell}$ is calculated for the entire $\ell\ell$ sample.  To estimate $\kappa$ and $A\times\epsilon_{ID}$,  we use inclusive $Z/\gamma^{*}\rightarrow \ell\ell$ 
events generated with the {\sc pythia}~\cite{pythia}
generator with final-state radiation simulated using {\sc photos}~\cite{photos}
and the CTEQ6.1L~\cite{cteq61} PDF set.  
Because {\sc pythia} is a leading-order (LO) generator and does not 
reproduce the observed $p_T^Z$ spectrum in data, generated events 
are weighted to reflect the $p_T^Z$ distribution observed in Ref.~\cite{zpt}.
Events are then traced through the D0  detector using a simulation 
based on {\sc geant} \cite{geant}.  
Data events from random beam crossings are overlaid 
on the simulated interactions to reproduce the effects of 
multiple $p\bar{p}$ interactions and detector noise. 
Simulated interactions that take into account the 
observed differences between data and simulation are reweighted, e.g., 
the $z$ coordinate of the vertex, instantaneous luminosity, trigger efficiency, 
lepton identification (ID) efficiency, photon ID efficiency, 
and resolution effects.  Here, the factor $(A\times\epsilon_{ID})_{\ell\ell}$ has values of 0.15 (0.17) in the electron channel (muon channel).  When no constraints on $M_{\ell\ell\gamma}$ are applied, the factor $\kappa$ has average values of $0.83 \pm 0.01$ (stat.)~and $0.85 \pm 0.01$ (stat.)~for the electron and muon channels, respectively, and the average value of $(A\times\epsilon_{ID})_{\ell\ell\gamma}$ is 0.12 for both the electron and muon channels.  Values for $(A\times\epsilon_{ID})_{\ell\ell\gamma}$ and $\kappa$ are similar for the subsample requiring $M_{\ell\ell\gamma}>110$ GeV/$c^2$.  

To account for systematic uncertainty on the migration into the sample from generated events with $p_T^\gamma <$ 10 GeV/$c$, 
we conservatively vary the number of events produced outside the generator-level requirements
in the {\sc pythia} simulation by $\pm$20\%, found as an upper estimate in studies of photon energy resolution in this kinematic regime, to measure the effect on the final cross section measurement.  
We find that the effect introduces a 1.5\% systematic uncertainty 
on the total cross section. 
The dominant uncertainty corresponding to the calculation of 
$A\times\epsilon_{ID}$ is due to choice of the PDF set. 
There are 20 free parameters in the CTEQ6.1L parametrization of the PDF that reflect fits to data from previous experiments.  The uncertainties 
on acceptance and efficiencies due to the PDF parametrization are estimated using the CTEQ6.1M PDF uncertainties, following Ref.~\cite{PDFerrs}.  We find a total PDF uncertainty 
of 3.5\%, dominated by the uncertainty on the acceptance-correction to the full geometrical lepton acceptance.  
The photon ID efficiency is determined from a simulated sample of photons and is estimated to 
have an uncertainty of 10\% for $p_T^\gamma<$ 15 GeV/$c$ and 3\% 
for $p_T^\gamma>$ 15 GeV/$c$.

To reduce the contribution of FSR in the data samples, we calculate the cross section with and without the $M_{\ell\ell\gamma}>110$ GeV/$c^2$ requirement. 
To combine the electron and muon channels, we utilize the method in Ref.~\cite{BLUE}, which averages the results of measurements with correlated systematic uncertainties.  We assume the  PDF and photon ID efficiency uncertainties to be 100\% correlated between the two channels.  The total cross section results can be found in Tables \ref{xsect} and \ref{xsect_llg110}.   The measurements are consistent with the NLO {\sc mcfm}  \cite{MCFM}
 prediction using CTEQ6.6 PDF set \cite{CTEQ66} and the renormalization and factorization scales evaluated 
at the mass of the $W$ boson, $M_W=80$ GeV/$c^2$.  The PDF uncertainties associated with the SM prediction are evaluated following Ref.~\cite{PDFerrs}.  We reevaluate the values for the $p_T^\gamma$ spectrum calculated by NLO {\sc mcfm} with the renormalization 
and factorization scales set to 160 GeV/$c^2$ and again at 40 GeV/$c^2$ 
and use these as estimates of the theoretical uncertainty of 1 standard deviation relative to the central NLO {\sc mcfm} value.

\begin{table}[h!]
\begin{ruledtabular}
\begin{center}
\caption{\label{xsect} Summary of the total cross-section measurements, when no $M_{\ell\ell\gamma}$ requirement is applied, for individual channels, combined channels, and the NLO {\sc mcfm} calculation with associated PDF and scale uncertainties. }
  \begin{tabular}{   l l l }
    & & \multicolumn{1}{c}{$\sigma_{Z\gamma} \times \mathcal{B}$ [fb]} \\ \hline 
    \multicolumn{2}{l}{$ee\gamma$ data} & 1026 $\pm$ 62 (stat.) $\pm$ 60 (syst.) \\ 
     \multicolumn{2}{l}{$\mu\mu\gamma$ data} & 1158 $\pm$ 53 (stat.) $\pm$ 70 (syst.) \\ \hline
     \multicolumn{2}{l}{$\ell\ell\gamma$ combined data} & 1089 $\pm$ 40 (stat.) $\pm$ 65 (syst.) \\ \hline 
    \multicolumn{2}{l}{NLO {\sc mcfm}} & 1096 $\pm$ 34 (PDF) $^{+2}_{-4}$ (scale) 
\end{tabular}

\end{center}
\end{ruledtabular}
\end{table}

\begin{table}[h!]
\begin{ruledtabular}
\begin{center}
\caption{\label{xsect_llg110} Summary of the total cross-section measurements, with the $M_{\ell\ell\gamma}>110$  GeV/$c^2$ requirement, for individual channels, combined channels, and the NLO {\sc mcfm} calculation with associated PDF and scale uncertainties. }
  \begin{tabular}{  l l l }
     & & \multicolumn{1}{c}{$\sigma_{Z\gamma} \times \mathcal{B}$ [fb]} \\ \hline 
    \multicolumn{2}{l}{$ee\gamma$ data} & 281 $\pm$ 17 (stat.) $\pm$ 11 (syst.) \\ 
    \multicolumn{2}{l}{$\mu\mu\gamma$ data} & 306 $\pm$ 28 (stat.) $\pm$ 11 (syst.) \\ \hline 
     \multicolumn{2}{l}{$\ell\ell\gamma$ combined data} & 288 $\pm$ 15 (stat.) $\pm$ 11 (syst.) \\ \hline 
     \multicolumn{2}{l}{NLO {\sc mcfm}} & 294 $\pm$ 10 (PDF) $^{+1}_{-2}$ (scale) \\ 
\end{tabular}
\end{center}
\end{ruledtabular}
\end{table}

\subsection{Differential cross section $d\sigma/dp_T^\gamma$}
\begin{figure}[htbp]
\includegraphics[width=0.45\textwidth]{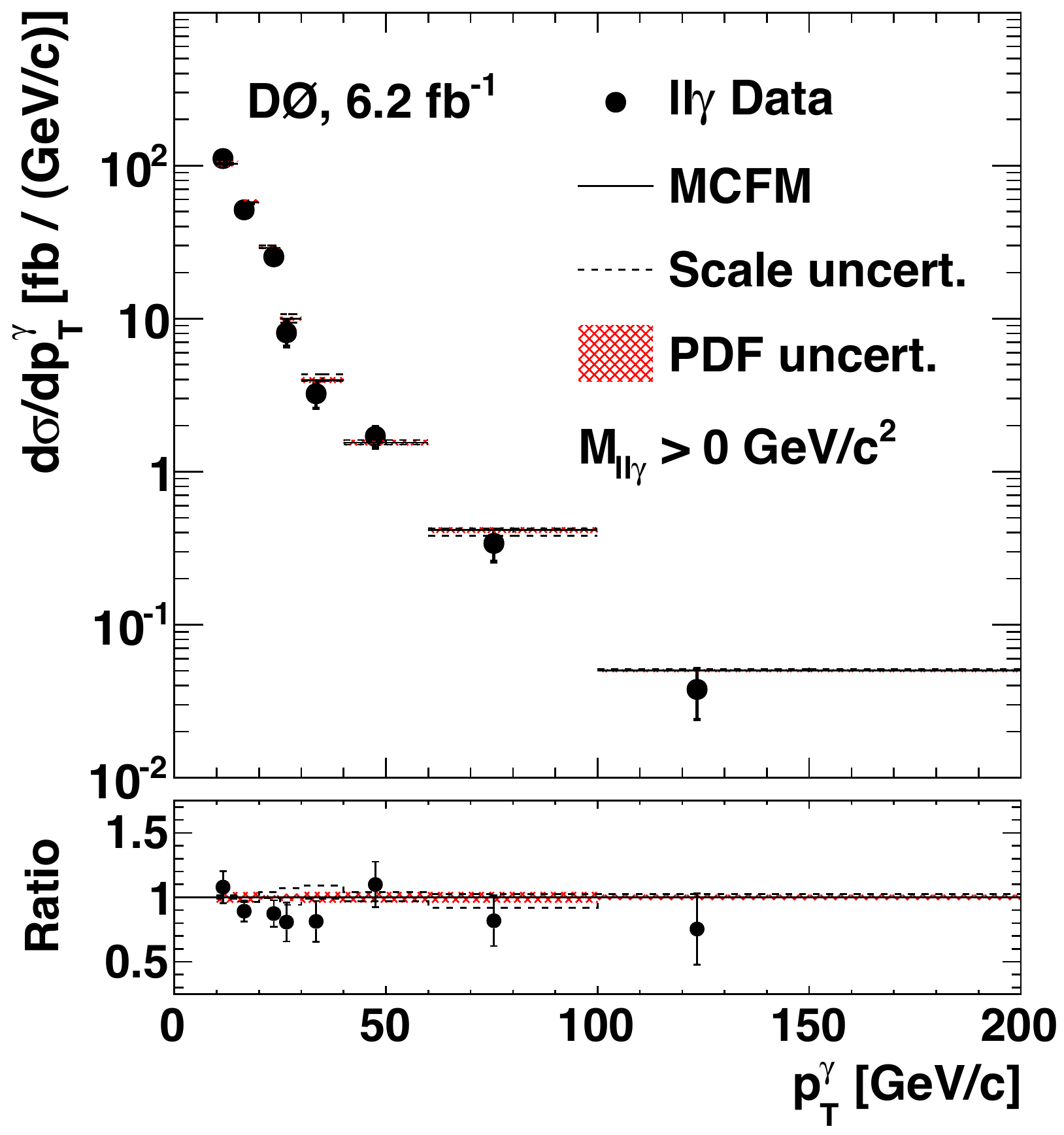}
\caption{Unfolded $d\sigma/dp_T^\gamma$ distribution with no $M_{\ell\ell\gamma}$ requirement for combined electron and muon data compared to the NLO {\sc mcfm} prediction. }
\label{unfolded}
\end{figure}

\begin{figure}[htbp]
\includegraphics[width=0.45\textwidth]{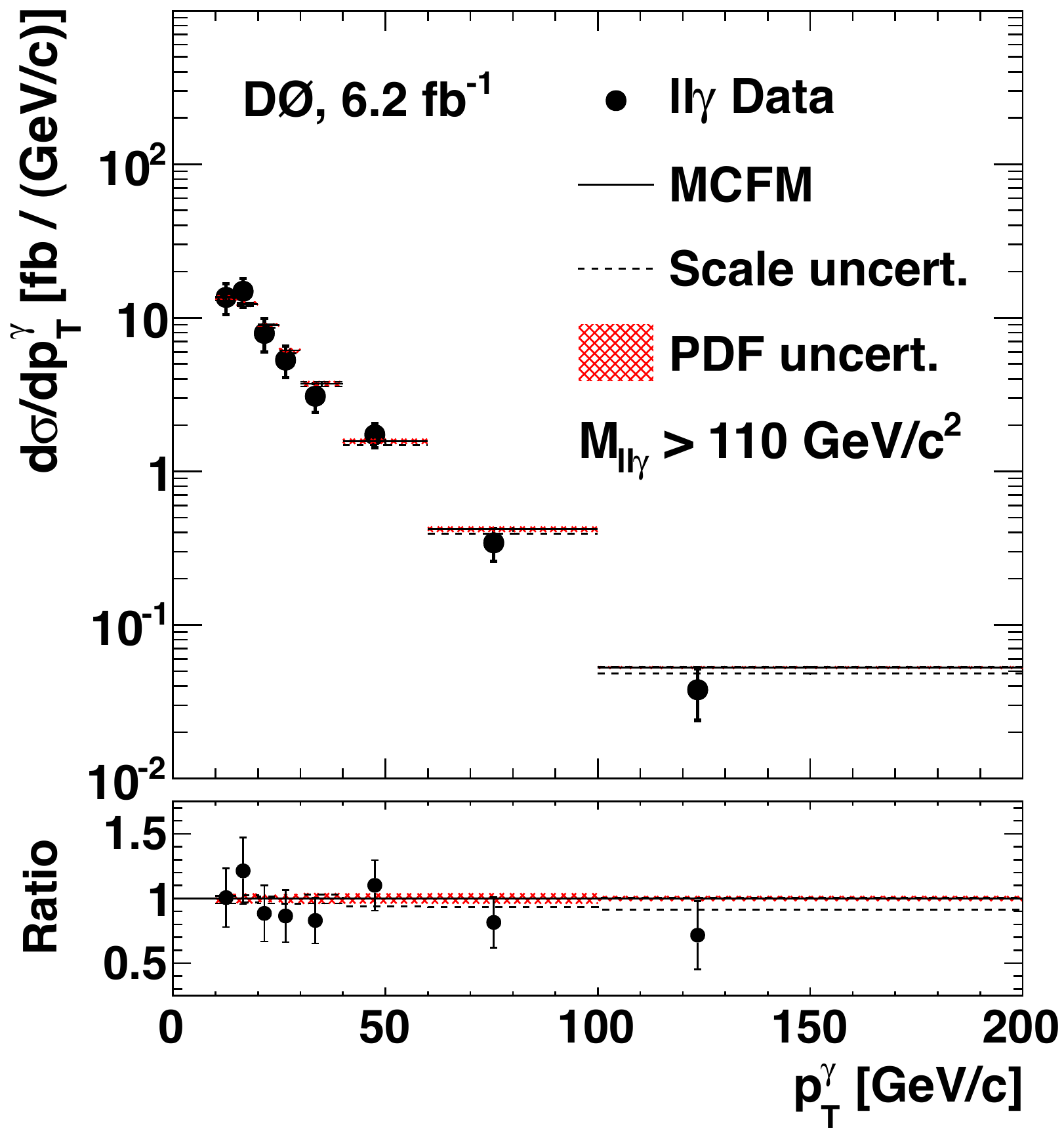}
\caption{Unfolded $d\sigma/dp_T^\gamma$ distribution with $M_{\ell\ell\gamma}>110$ GeV/$c^2$ for combined electron and muon data compared with the NLO {\sc mcfm} prediction. }
\label{unfolded_llg110}
\end{figure}

We use the matrix inversion technique \cite{unfolding} to unfold the experimental resolution and extract $d(\sigma_{Z\gamma}\times\mathcal{B})/dp_T^\gamma$ ($d\sigma/dp_T^\gamma$ henceforth), the differential cross section for $Z\gamma\rightarrow \ell\ell\gamma$, as a function of the true $p_T^\gamma$.  
The elements of the smearing matrix between true and reconstructed $p_T^\gamma$ bins are estimated using the full simulation of the
   detector response on a sample of $Z\gamma$ events generated
   using {\sc pythia}.  Then, the matrix is inverted to 
obtain the unsmeared spectrum.  We confirm that the unfolding procedure introduces a negligible bias.  
Following Ref.~\cite{bincentering}, the position of the data points are plotted in Figs.~\ref{unfolded} and \ref{unfolded_llg110} at the value of $p_T^\gamma$  where
the cross section equals the average value for that bin.  The theoretical uncertainties associated with the choice of PDF and the renormalization and factorization scales are determined analogously to the theoretical prediction for the total production cross section.  
  The combined differential cross sections $d\sigma/dp_T^\gamma$ are shown in  Figs.~\ref{unfolded} and \ref{unfolded_llg110} for no $M_{\ell\ell\gamma}$ requirement and $M_{\ell\ell\gamma}>110$ GeV/$c^2$, respectively.  The values associated with Figs.~\ref{unfolded} and \ref{unfolded_llg110} are given in Tables \ref{unfoldedvals} and \ref{unfoldedvals_llg110}.

\begin{table*}
\begin{ruledtabular}
\caption{\label{unfoldedvals} Summary of the unfolded differential cross section $d\sigma/dp_T^\gamma$, when no $M_{\ell\ell\gamma}$ requirement is applied, and NLO {\sc mcfm} predictions with PDF and scale uncertainties}
  \begin{tabular}{ D{,}{-}{-1}   c  l  l  } 
     \multicolumn{2}{c}{ }  & \multicolumn{1}{c}{$\ell\ell\gamma$ combined data} & \multicolumn{1}{c}{NLO {\sc mcfm}} \\ \hline
   \multicolumn{1}{c}{$p_T^\gamma$ bin }  & $p_T^\gamma$ center & \multicolumn{2}{c}{$d\sigma/d  p_T^\gamma $ }    \\ 
    \multicolumn{1}{c}{\mbox{[GeV/$c$]}} &  [GeV/$c$] & \multicolumn{2}{c}{ [fb/(GeV/$c$)]}     \\ \hline
   10,15 & 12.4 	&	111.14 $\pm$ 4.40 (stat.)~$\pm$ 11.99 (syst.)  	& 104.02 $\pm$ 4.10 (PDF)   $^{+1.4}_{-1.2}$ (scale)  		\\ 
   15 , 20 & 17.2	&	51.41 $\pm$ 3.83 (stat.)~$\pm$ 2.65 (syst.) 		&  57.13 $\pm$ 2.23 (PDF)  $^{+1.3}_{-1.8}$ (scale)		\\ 
   20 , 25 & 22.5	&	25.34 $\pm$ 2.74 (stat.)~$\pm$ 1.13	 (syst.) 	& 28.77 $\pm$ 0.43 (PDF)  $^{+1.1}_{-0.7}$ (scale)   		\\ 
   25 , 30 & 27.5	&	8.08 $\pm$ 1.45 (stat.)~$\pm$ 0.40	(syst.) 	& 10.16 $\pm$ 0.26 (PDF)  $^{+0.7}_{-0.5}$ (scale)    		\\ 
   30 , 40 & 34.4	&	3.23 $\pm$ 0.60 (stat.)~$\pm$ 0.17	(syst.) 	& 4.15 $\pm$ 0.16 (PDF)    $^{+0.34}_{-0.19}$ (scale) 			\\ 
   40 , 60 & 48.5	&	1.70 $\pm$ 0.26 (stat.)~$\pm$ 0.088	 (syst.) 	& 1.60 $\pm$ 0.061 (PDF)  $^{+0.008}_{-0.010}$ (scale)   		\\ 
   60 , 100 & 76.5	&	0.34 $\pm$ 0.079 (stat.)~$\pm$ 0.018 (syst.) 	& 0.42 $\pm$ 0.017 (PDF)    $^{+0.028}_{-0.028}$ (scale)  		\\ 
   100 , 200 & 124.5	&	0.038 $\pm$ 0.014 (stat.)~$\pm$ 0.002 (syst.) 	& 0.052 $\pm$ 0.001 (PDF)    $^{+0.003}_{-0.001}$ (scale) 		\\  
  \end{tabular}
\end{ruledtabular}
\end{table*}

\begin{table*}
\begin{ruledtabular}
\caption{\label{unfoldedvals_llg110} Summary of the unfolded differential cross section $d\sigma/dp_T^\gamma$, with the $M_{\ell\ell\gamma}>110$ GeV/$c^2$ requirement, and NLO {\sc mcfm} predictions with PDF and scale uncertainties .}
 \begin{tabular}{ D{,}{-}{-1}   c  l   l  } 
     \multicolumn{2}{c}{ }  & \multicolumn{1}{c}{$\ell\ell\gamma$ combined data} & \multicolumn{1}{c}{NLO {\sc mcfm}}\\ \hline
   \multicolumn{1}{c}{$p_T^\gamma$ bin }  & $p_T^\gamma$ center & \multicolumn{2}{c}{$d\sigma/d  p_T^\gamma $ }    \\ 
    \multicolumn{1}{c}{\mbox{[GeV/$c$]}} &  [GeV/$c$] & \multicolumn{2}{c}{ [fb/(GeV/$c$)]}     \\ \hline
   10 , 15 & 13.7 	&	13.57 $\pm$ 1.87 (stat.)~$\pm$ 2.43 (syst.)  		& 13.48 $\pm$ 0.48 (PDF) $^{+0.25}_{-0.51}$ (scale)    		\\ 
   15 , 20 & 17.2	&	14.87 $\pm$ 2.17 (stat.)~$\pm$ 2.30 (syst.) 		& 12.25 $\pm$ 0.47 (PDF)  $^{+0.29}_{-0.36}$ (scale)    		\\ 
   20 , 25 & 22.0	&	7.91 $\pm$ 1.76 (stat.)~$\pm$ 0.81	(syst.) 	& 8.94 $\pm$ 0.25 (PDF) $^{+0.13}_{-0.35}$ (scale)  		\\ 
   25 , 30 & 27.4	&	5.30 $\pm$ 1.15 (stat.)~$\pm$ 0.44	(syst.) 	& 6.13  $\pm$ 0.21(PDF) $^{+0.016}_{-0.25}$ (scale)  	\\ 
   30 , 40 & 34.5	&	3.08 $\pm$ 0.57 (stat.)~$\pm$ 0.33	(syst.) 	& 3.71  $\pm$ 0.15  (PDF) $^{+0.012}_{-0.14}$ (scale)  			\\ 
   40 , 60 & 48.6	&	1.73 $\pm$ 0.26 (stat.)~$\pm$ 0.17	(syst.) 	& 1.57 $\pm$ 0.061 (PDF) $^{+0.004}_{-0.094}$ (scale)    	\\ 
   60 , 100 & 76.5	&	0.34 $\pm$ 0.079 (stat.)~$\pm$ 0.019 (syst.) 		& 0.42 $\pm$ 0.017 (PDF) $^{+0.028}_{-0.028}$ (scale)    		\\ 
   100 , 200 & 124.5	&	0.038 $\pm$ 0.014 (stat.)~$\pm$ 0.002 (syst.) 	& 0.052 $\pm$ 0.001  (PDF) $^{+0.003}_{-0.001}$ (scale)  		\\  
  \end{tabular}
\end{ruledtabular}
\end{table*}

\begin{table}[h!]
\begin{ruledtabular}
\begin{center}
\caption{\label{1Dlimits} Summary of the 1D limits on the $ZZ\gamma$ and $Z\gamma\gamma$ coupling parameters at the 95\% C.L.  }
  \begin{tabular}{   l  c  c  c } 
  & &  &  $\ell\ell\gamma$ 7.2 fb$^{-1}$ \\ 
  & \multicolumn{2}{c}{$\ell\ell\gamma$ 6.2 fb$^{-1}$} & $\nu\nu\gamma$ 3.6 fb$^{-1}$ \\
  & $\Lambda=1.2$ TeV & $\Lambda=1.5$ TeV & $\Lambda = 1.5$ TeV \\ \hline
  $|h^Z_{03}|<$ & 0.050 & 0.041 & 0.026\\ 
  $|h^Z_{04}|<$ & 0.0033 & 0.0023 & 0.0013\\ \hline
   $|h^\gamma_{03}|<$ & 0.052 & 0.044 & 0.027\\ 
  $|h^\gamma_{04}|<$ & 0.0034 & 0.0023 & 0.0014\\ 
\end{tabular}
\end{center}
\end{ruledtabular}
\end{table}

\section{Limits on Anomalous Couplings}
\begin{figure}[h!]
\begin{minipage}[b]{0.9\linewidth}
\centering
\includegraphics[scale=0.45]{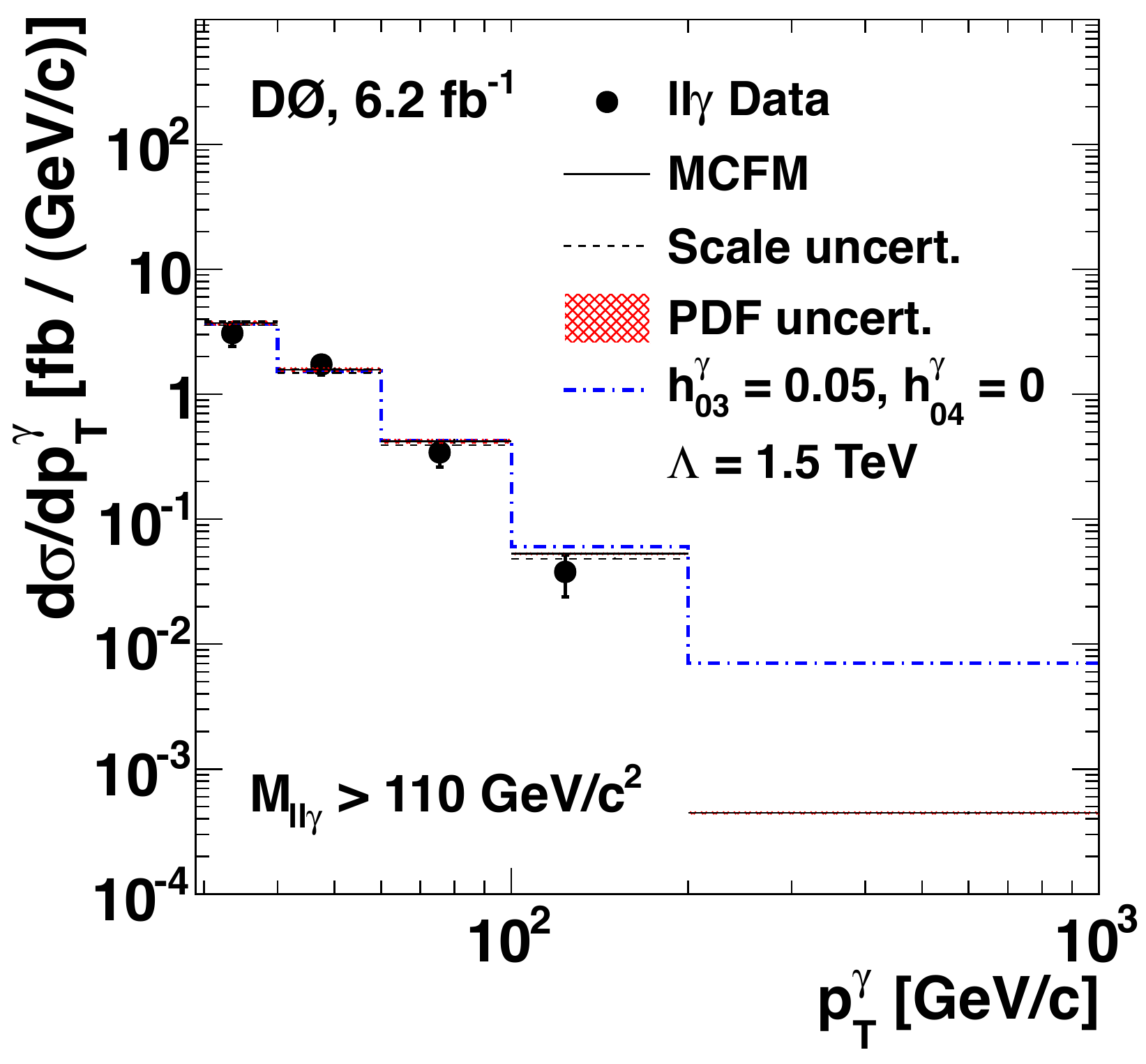}
\caption{\label{ACplot} The SM prediction and anomalous $Z\gamma$ coupling production compared with the unfolded $d\sigma/dp_T^\gamma$ for combined muon and electron channels for $p_T^\gamma > 30$ GeV/$c$ and $M_{\ell\ell\gamma}>110$ GeV/$c^2$.}
\end{minipage}
\end{figure}

\begin{figure}
\subfigure[]{\label{fig:ISR1}\includegraphics[scale=.45]{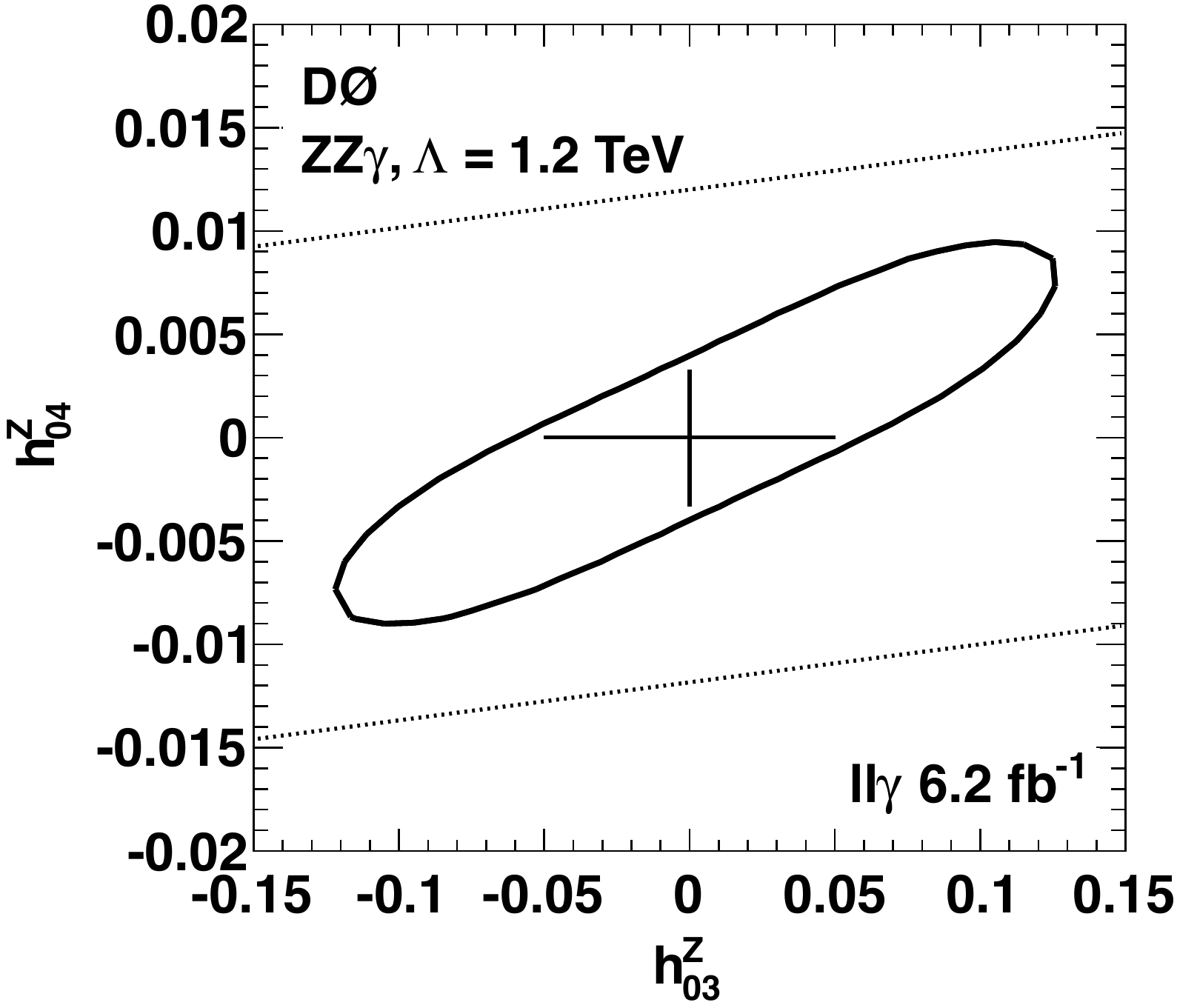}}
\subfigure[]{\label{fig:ISR1}\includegraphics[scale=.45]{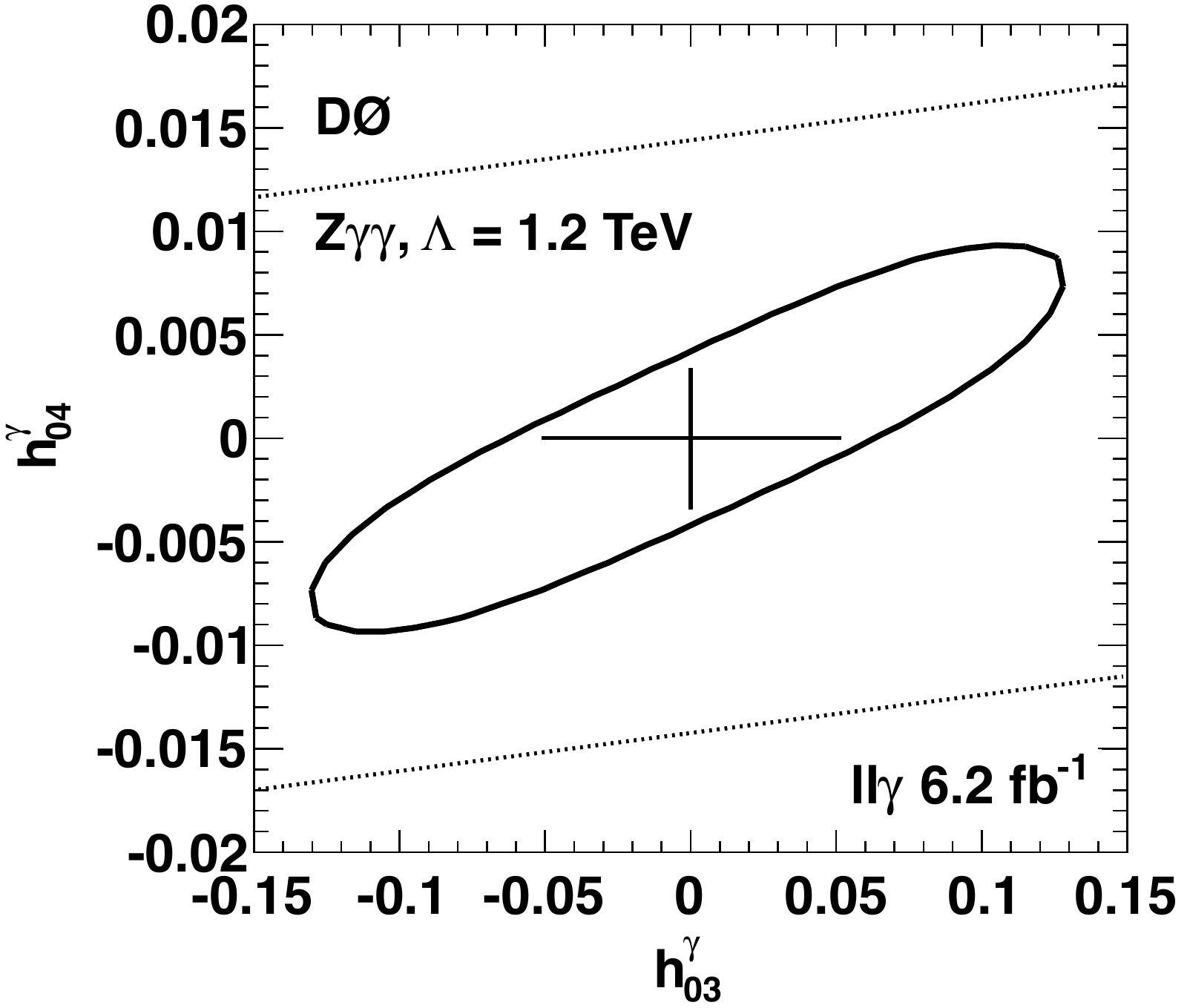}}
\caption{The 2D (contour) and 1D (cross) limits on the anomalous coupling parameters for (a) $ZZ\gamma$ and (b)  $Z\gamma\gamma$ vertices at the 95\% C.L. for $\Lambda=1.2$ TeV.  Limits on $S$-matrix unitarity are represented by the dotted lines. }
\label{2D_1200}
\end{figure}

\begin{figure}
\subfigure[]{\label{fig:ISR1}\includegraphics[scale=.45]{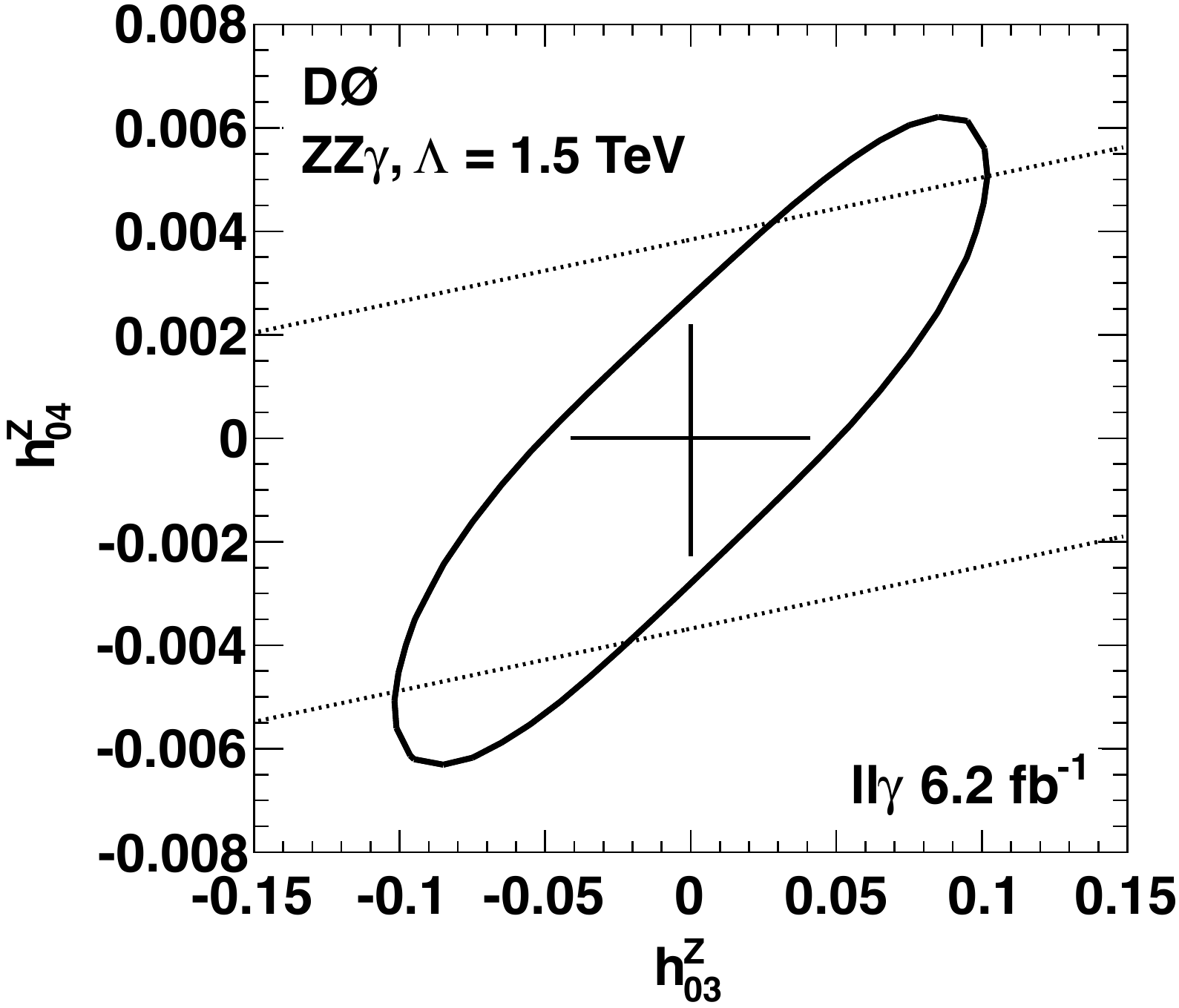}}
\subfigure[]{\label{fig:ISR1}\includegraphics[scale=.45]{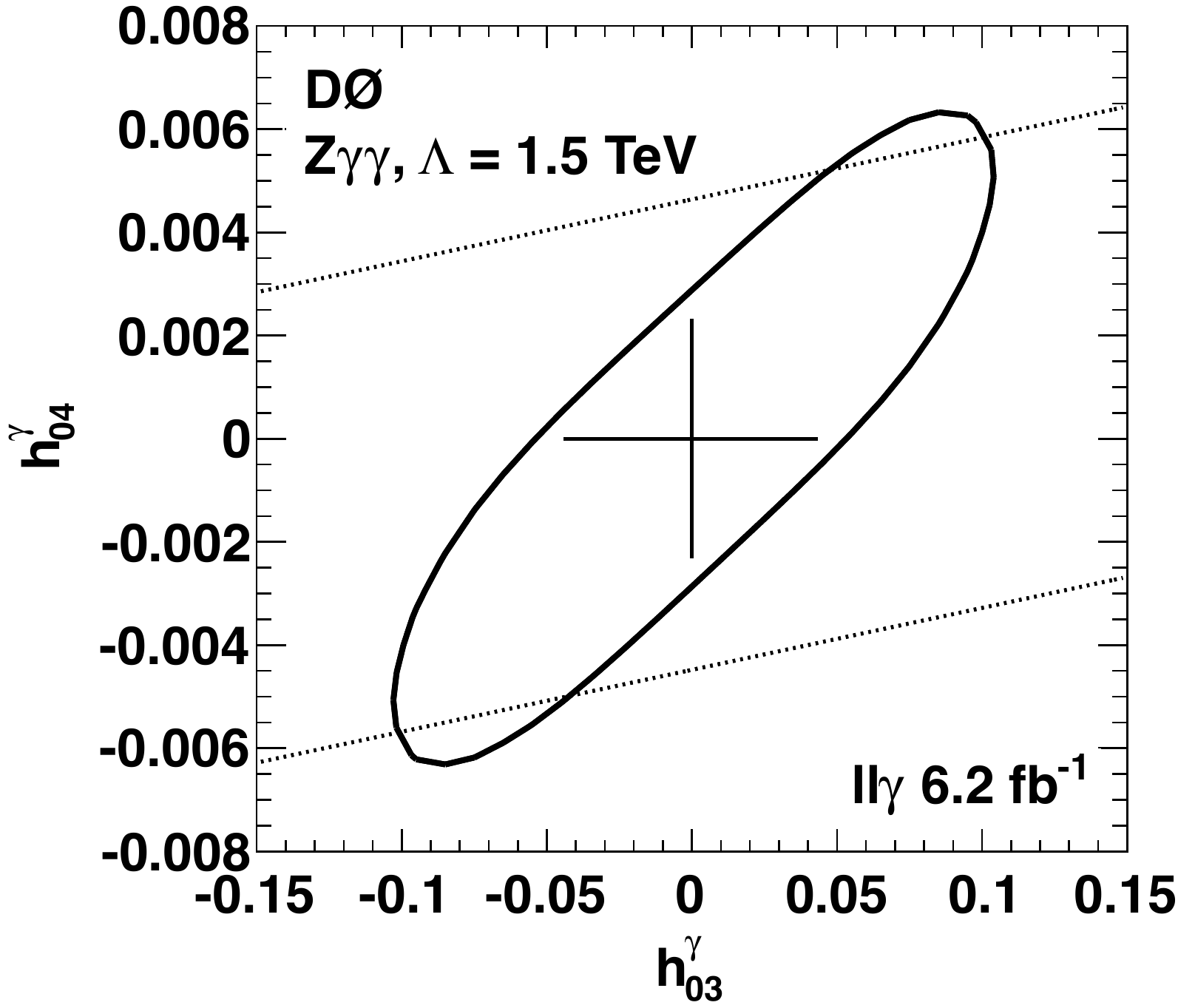}}
\caption{The 2D (contour) and 1D (cross) limits on the anomalous parameters for (a) $ZZ\gamma$ and (b)  $Z\gamma\gamma$ vertices at the 95\% C.L. for $\Lambda=1.5$ TeV.  Limits on $S$-matrix unitarity are represented by the dotted lines. }
\label{2D_1500}
\end{figure}

To set limits on anomalous trilinear gauge boson couplings, 
we generate $Z\gamma$ events for different values of the anomalous couplings using the NLO Monte Carlo generator of Ref.~\cite{baur}. SM Drell-Yan production is included by reweighting the $p_T^\gamma$ spectrum to {\sc mcfm} for vanishing anomalous couplings.
As shown in Fig.~\ref{ACplot}, anomalous $Z\gamma$ couplings would contribute to an excess of high-energy photons as compared to the SM prediction. We apply the following generator-level requirements:
$M_{\ell\ell}>60$ GeV/$c^2$, $\Delta \mathcal{R}_{\ell\gamma} > 0.7$, $p_T^\gamma>10$ GeV/$c$, 
$|\eta^\gamma|<1$,  
and $M_{\ell\ell\gamma}>$110 GeV/$c^2$, generate $p_T^\gamma$ templates as a function of the anomalous couplings, and use the known acceptance and resolution functions to fold the predicted generator-level distribution into 
a reconstruction-level distribution for $p_T^\gamma$. 
Using Poisson statistics for $p_T^\gamma > 30$ GeV/$c$, we define a likelihood function to 
compare the combined electron and muon channels 
with a predicted distribution for given values of anomalous couplings.
In the absence of any significant deviation from
   the SM prediction, we set one-dimensional (1D) and two-dimensional (2D) limits 
on the anomalous coupling parameter values at the 95\% C.L.  A combined log-likelihood function using all data is defined by the sum of the individual log-likelihood functions of the electron and muon channels.  We include the effect of systematic uncertainties associated with transforming a Monte Carlo $p_T^\gamma$ template from the generator-level into a reconstructed distribution and find that these uncertainties contribute to the value of the calculated limits on the order of 1\%.   
We generate a $10 \times 10$ grid of templates for the $p_T^\gamma$ distribution as a function of $h^V_{03}$ and $h^V_{04}$ for $|h^V_{03}|<0.1$ and $|h^V_{04}|<0.01$, while setting all other coupling parameters to zero, and the limits are derived by varying about the maxima of the log-likelihood functions for the 95\% C.L \cite{PDGstats}.  Results for the 1D limits for $\Lambda = 1.2$ TeV and 1.5 TeV are shown in Table \ref{1Dlimits}.  The 1D and 2D limits on the anomalous coupling parameters are shown in Figs.~\ref{2D_1200} and \ref{2D_1500}, utilizing the electron and muon channels.  In these figures, the dotted lines represent the theoretical limits on the anomalous coupling values, beyond which $S$-matrix unitarity is violated.  Because the $h_{04}^V$ parameters come from dimension-eight operators, the limits are more constrained than those of $h_{03}^V$ couplings, which are dimension-six.

\begin{figure}[h!]
\begin{minipage}[b]{1.0\linewidth}
\centering
\subfigure[]{\label{fig:ISR1}\includegraphics[scale=.45]{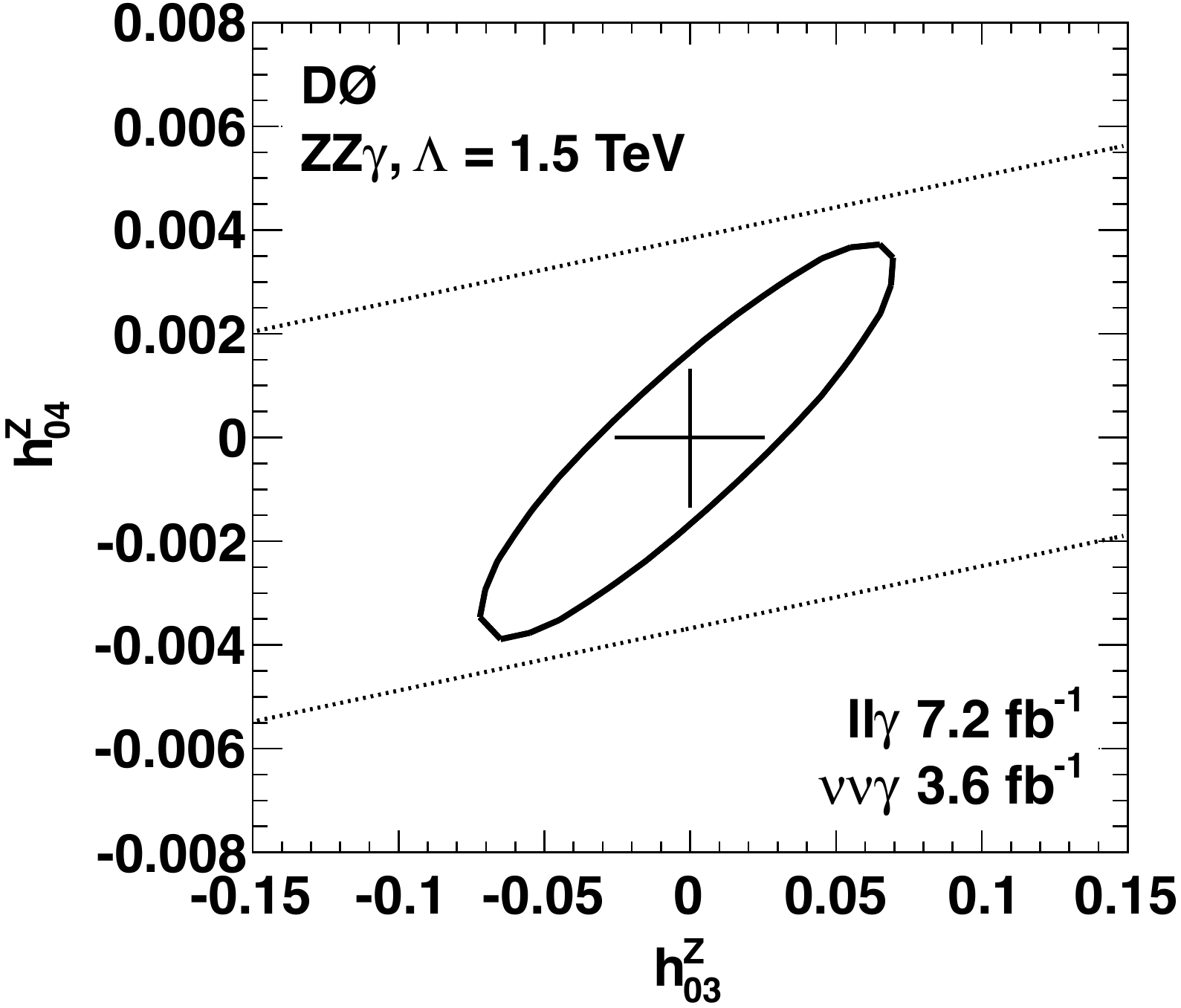}}
\subfigure[]{\label{fig:ISR1}\includegraphics[scale=.45]{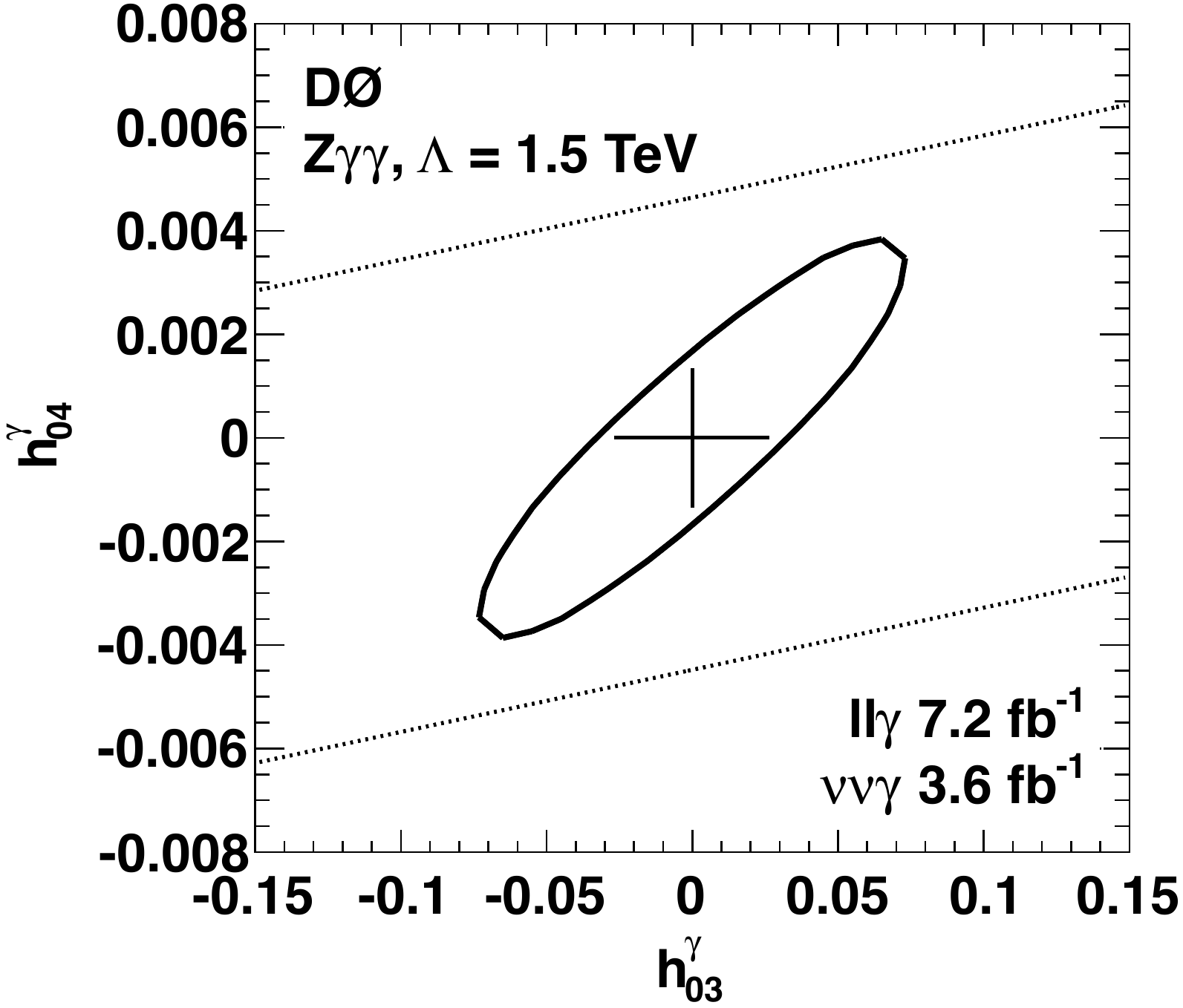}}
\caption{\label{2D_1500_wprev} The 2D (contour) and 1D (cross) limits on coupling parameters for (a) $ZZ\gamma$ and (b)  $Z\gamma\gamma$ vertices at the 95\% C.L. for $\Lambda=1.5$ TeV.  Limits on $S$-matrix unitarity are represented by the dotted lines. }
\end{minipage}
\end{figure}

We combine these results with those of a previous D0 $Z\gamma$ analysis \cite{d0prev} .  In that analysis, the 1D and 2D limits on the anomalous couplings parameter were calculated using a data sample corresponding to an integrated luminosity of 1 fb$^{-1}$ of data collected between October~2002 and Feburary~2006 (3.6 fb$^{-1}$ of data collected between October~2002 and September~2008) in the $ee\gamma$ and $\mu\mu\gamma$ channels ($\nu\nu\gamma$ channel), for $\Lambda = 1.5$ TeV.  Results can be found in Fig.~\ref{2D_1500_wprev} and Table \ref{1Dlimits}.

\section{Conclusions}

We have measured the differential and total cross sections 
for $Z\gamma\rightarrow \ell\ell\gamma$ production in $p\bar{p}$ collisions using the D0 detector at the Tevatron Collider with and without a $M_{\ell\ell\gamma}>110$ GeV/$c^2$ requirement.  
Both the total production cross sections and differential cross sections $d\sigma/dp_T^\gamma$ are consistent with the SM at NLO predicted 
by {\sc mcfm} \cite{MCFM}.  We observe no deviation from SM predictions and place 1D and 2D limits on the $CP$-even anomalous $Z\gamma$ couplings for $\Lambda=1.2$ and 1.5 TeV.  When combining with the previous D0 analyses, the limits are comparable to those found in the most recent CDF result \cite{zgammaprev5}, which uses $\approx 5$ fb$^{-1}$ in the $\ell\ell\gamma$ and $\nu\bar{\nu}\gamma$ channels and $\Lambda=1.5$ TeV.  Our results include the first unfolded photon differential cross section $d\sigma/dp^\gamma_T$, as well as the most precise measurement of the total production cross section of $Z\gamma\rightarrow \ell\ell\gamma$.  

\begin{acknowledgements}
%
We thank the staffs at Fermilab and collaborating institutions,
and acknowledge support from the
DOE and NSF (USA);
CEA and CNRS/IN2P3 (France);
FASI, Rosatom and RFBR (Russia);
CNPq, FAPERJ, FAPESP and FUNDUNESP (Brazil);
DAE and DST (India);
Colciencias (Colombia);
CONACyT (Mexico);
KRF and KOSEF (Korea);
CONICET and UBACyT (Argentina);
FOM (The Netherlands);
STFC and the Royal Society (United Kingdom);
MSMT and GACR (Czech Republic);
CRC Program and NSERC (Canada);
BMBF and DFG (Germany);
SFI (Ireland);
The Swedish Research Council (Sweden);
and
CAS and CNSF (China).

\end{acknowledgements}


\begin{thebibliography}{99}
\bibitem{d0_coordinate}
Pseudorapidity is defined as $\eta=-\ln [\tan(\theta/2)]$, where
$\theta$ is the polar angle relative to the proton beam direction.
$\phi$ is defined to be the azimuthal angle in the plane transverse to
the proton beam direction.
\end{thebibliography}
\end{document}